**Creation of super-high-flux photo-neutrons and gamma-rays > 8 MeV using a petawatt laser to irradiate high-Z solid targets**

E.Liang[1], W.Lo[2], B.Cage[1], E. Fang[1], S.Arora[1], K.Q.Zheng[1], H.Quvedo[2], S.A.Bruce[2], M. Spinks[2], E. Medina[2], A. Helal[2], T. Ditmire[2]

1 Rice University, Houston, TX 77005
2 University of Texas at Austin, Austin, TX 78712

**Abstract**

We report the creation of super-high-flux gamma-rays with energy >8 MeV and photo-neutrons via the ($\gamma$,n) reaction near giant dipole resonance energies (8 - 20 MeV), using the ~130 J Texas Petawatt laser to irradiate high-Z (Au, Pt, Re, W) targets of mm - cm thickness, at laser intensities up to ~$5x10^{21}$W/cm$^2$. We detected up to ~ several x $10^{12}$ gamma-rays > 8 MeV (~3% of incident laser energy) and ~ $10^{10}$ photo-neutrons per shot. Due to the short pulse and narrow gamma-ray cone (~17$^o$ half-width) around laser forward, the peak emergent gamma-ray flux >8 MeV reached ~$10^{27}$ gammas/cm$^2$/sec, and the peak emergent neutron flux reached ~$10^{20}$ neutrons/cm$^2$/sec. Such intense gamma-ray and neutron fluxes are many orders of magnitudes higher than those achievalbe with reactors or accelerators and among the highest achieved for short-pulse laser experiments > 100 J. They will facilitate the study of nuclear reactions requiring super-high-flux of gamma-rays or neutrons, such as the creation of r-process elements. These results may also have far-reaching applications for nuclear energy, such as the transmutation of nuclear waste, isotope production and inertial fusion.

## 1. INTRODUCTION

Modern ultra-intense short-pulse lasers are efficient accelerators of relativistic electrons (Gibbon 2005). When these relativistic electrons interact with high-Z solid targets, they produce copious amounts of gamma-rays, which in turn create electron-positron pairs (Liang et al 1998, Chen H. et al 2009, Liang et al 2015) and photo-neutrons via ($\gamma$,Nn) reactions (Makwana et al 2017). While many different acceleration mechanisms have been proposed for the relativistic "hot" electrons (Gibbon 2005, Levy et al 2014), it is generally assumed that the gamma-rays are mainly emitted by electron bremsstrahlung against target ions (Rybicki & Lightman 1979, Chen et al 2009, Henderson et al 2014). The bremsstrahlung spectrum falls exponentially with photon energy, with the exponential constant determined by the average energy of primary hot electrons and secondary electrons and positrons (Henderson 2015). The bremsstrahlung emission from laser-solid interactions has been well characterized below ~ 6 MeV as ~exponential by many groups using filter-stack spectrometers (FSS. Chen et al 2009, Henderson 2015). However, the FSS cannot determine the gamma-ray spectrum > 6 MeV (Liang et al 2022). While other schemes (e.g. nuclear activation thresholds (Leeman et al 2001), forward Compton scattering (Morgan et al 1991, Kojima et al 2014) ) have been used in some laser experiments, they are strongly model-dependent. Most laser-created gamma-ray spectra > 6 MeV are based on few-channel forward-folding using a simple exponential model (Belm et al 2018, Gunther et al 2022).

A radically new type of high-resolution gamma-ray spectrometer called scintillation attenuation spectrometer (SAS) was recently developed by our group in collaboration with MD Anderson Cancer Center medical imaging group, based on 2D imaging of scintillation light from a highly pixelated LYSO crystal matrix with mm-sized pixels (Liang et al 2022). The SAS can measure gamma-ray spectrum from 0.25 MeV - 50 MeV with 0.5 MeV resolution. From 2016-2018, we obtained ~180 gamma-ray spectra from two SAS viewing different angles, from mm-cm thick Au and Pt targets irradiated by the ~130 J Texas Petawatt laser (TPW) in Austin Texas, at intensities ~ $10^{21}$ W.cm$^2$. However, a robust inversion algorithm to extract model-independent gamma-ray spectra from SAS images was only completed, calibrated and validated in 2021 (Liang et al 2022). In 2022 we began systematic extraction of TPW gamma-ray spectra from archived SAS data. One unexpected discovery was the presence of a broad high-energy (> 8 MeV) bump-like feature sitting on top of the conventional exponential bremsstrahlung spectrum (Fig.1). This broad feature peaks at ~10-20 MeV and often contains more energy than the bremsstrahlung component. However, this gamma-ray "bump" is absent in spectra obtained at angles far from laser forward (LF), where only the exponential bremsstrahlung spectrum is observed (Liang et al 2022) For easy reference, hereafter we refer to the broad bump-like excess > 8 MeV as GRB, and the low-energy bremsstrahlung component as LEB. Fig.1 shows two SAS raw images and their corresponding gamma-ray spectra obtained using the inversion algorithm published in (Liang et al 2022). Fig1a shows the GRB and LEB components at LF, while Fig.1b shows that only the LEB component is present at 90$^o$ from LF.

Coincidentally, the cross-section for ($\gamma$,n) photo-neutron reactions of high-Z elements exhibits a strong peak spanning 8 - 20 MeV, corresponding to the giant dipole resonance (GDR, Makwana et al 2017, Kimura et al 2016)(cf. Fig.2). The strong overlap of the GDR peak with the GRB gamma-ray excess of Fig.1 motivated new TPW experiments in 2022 to study photonuclear reactions in the GDR regime. The goal was two-fold. We wanted to use photo-neutron yields to independently characterize the gamma-ray spectrum > 8 MeV. At the same time, we wanted to explore using TPW as a platform to create high-flux photo-neutrons and study photofission of actinides.

We completed 60 shots in 2022 with TPW irradiating mm - cm thick targets of Au, Pt, Re and W, with intensities up to ~5 x $10^{21}$ W.cm$^{-2}$. We achieved superhigh-fluxes of photon-neutrons and gamma-rays > 8 MeV. We detected up to several x $10^{12}$ gamma-rays > 8 MeV (~ 3% of laser energy) and ~$10^{10}$ photo-neutrons per shot. The highest emergent gamma-ray flux >8 MeV reached ~$10^{27}$gammas/cm$^2$/sec, and the highest emergent photo-neutron flux reached ~$10^{20}$ neutrons/cm$^2$/sec, which formally reaches the threshold for r-process element creation (Fig.3) (Kajimo et al 2019, Chen et al 2019, Tanaka et al 2020, Goriely & Pinedo 2015). The super-high gamma-ray flux should also facilitate the study of photo-fission reactions of actinides (Naik et al 2011, Pomme 1994).

## 2. EXPERIMENT SETUP AND DIAGNOSTICS

Fig.4 is a sketch of the TPW high-intensity target chamber (TC1, Martinez et al 2005) where all our experiments were carried out with sample placements of diagnostics. In 2022 we deployed one e+e- magnetic spectrometer (MS, 0.1-65 MeV) inside TC1, and two SAS's (0.25-50 MeV) outside the TC1 tank wall. A FSS (≤ 6 MeV) is co-located with each SAS. Typically, one SAS will sit inside the gamma-ray cone (~17$^o$ half-angle, Henderson et al 2014, Henderson 2015) near laser forward (LF) , and a second SAS will be ~90$^o$ from LF. Up to 15 gamma dosimeters (0.02-6 MeV) were attached to the outside tank wall, replaced once a day, around the forward gamma-ray cone to monitor the gamma dose angular distribution. Neutron bubble detectors were attached to the outside tank wall at various positions to measure the neutron dose. The positions of the MS, SAS and FSS can be moved from shot to shot as we perform the angle sweep of the e+e- pair and gamma-ray outputs. Fig. 5 shows typical TPW focal spot intensity and time profiles during our run. Table 1 summarizes the laser and target parameters of our 2022 run. Except for a few outliers, TPW pulse energy and intensity were stable during our 60-shot run. But the pulse duration for the first half of our run was longer than usual, leading to an average pulse length of ~ 160 fs instead of the normal average of 140 fs.

During the second half of our run we employed a special technique to enhance the gamma-ray output. We discovered that by sending in a low-energy OPA prepulse to create a micro-divot of ~ Rayleigh length in size before sending in the main pulse, the gamma-ray output can be increased by up to a factor of 2. This technique was used in all our shots in which photo-neutron outputs were measured.

Two types of laser targets were studied in 2022: disks and capsules, with total thickness ranging from 1 mm to 2 cm. Flat circular or square disks of thickness from 1 mm to 1 cm of pure Au, Pt, Re and W were used to study the Z-dependence of gamma-rays and e+e- pairs. Neutrons were measured only for Pt disks. Fig.6 shows the typical geometry of a capsule target, which consists of a cm-square front cover made of pure Re (thickness 1.8 mm to 1cm) acting as the gamma-ray "converter", attached to a cm-cube of W "catcher" with a cylindrical divot along the central axis. The capsule design is for later experiments in which potentially activatable material can be securely housed and contained inside the divot while safely shielded from the target chamber and target handlers. In this paper we only report the results from empty capsules such that the gamma-rays and neutrons are produced only by the Re converter and W catcher. Data from filled capsules remain to be processed and analyzed. We chose Re as the converter material because Re target makes the smallest crater among the four elements, due to its low thermal conductivity, high density, high melting and boiling points, and high heats of fusion and vaporization. Hence Re targets allow the maximum number of laser shots within a given target area.

## 3. RESULTS

Fig.1 shows typical 2022 gamma-ray spectra obtained from our two SAS spectrometers. Near LF (Fig.1a) we obtain gamma-ray spectra containing a broad "bump" peaking around 13 - 20 MeV, in addition to the exponential component consistent with bremsstrahlung emission . We also use the dosimeter data to independently normalize the amplitude of the SAS spectra. The broad GRB feature is absent in the SAS data at ~ 90$^o$ from LF (Fig.1b), and we detect only the exponential component consistent with bremsstrahlung emission (LEB).

When the GRB excess > 8 MeV was first discovered in our 2016-2018 SAS data, we had no independent data to confirm or disprove it. Hence in our 2022 TPW experiment, we adopted two new approaches to independent constrain and ascertain the SAS gamma-ray spectrum. First, since the (γ, n) reaction cross-section has a well-characterized GDR peak spanning 8 MeV - 18 MeV for all high-Z elements (Fig.2), we can use the photo-neutron yield to independently constrain the gamma-ray spectrum in this energy range. Second, the high-energy e+e- pairs created by gamma-rays > 8 MeV are much less attenuated in thick targets than low-energy pairs created by bremsstrahlung gamma-rays < 6 MeV. By measuring the emergent spectra of electrons and positrons from cm-thick targets we should be able to cross-check the gamma-ray spectrum > 8 MeV which created them because pairs created by bremsstrahlung photons should peak at only 1-2 MeV. In particular, for cm-thick targets, we expect the electron and positron spectra to resemble each other since any sheath electric field would be negligible, and most of the laser accelerated primary hot electrons would be absorbed. It turns out this was indeed the case. Both the photo-neutron data and e+e- pair data for cm-thick targets are consistent with the gamma-ray spectrum of Fig.1a, and completely inconsistent with hot electron bremsstrahlung as the sole source of gamma-rays.

In most of our 2022 shots, the energy of the GRB gamma-rays > 8 MeV exceeds the energy of the LEB component < 8 MeV and reaches up to ~ 3% of incident laser energy. For example in Fig.1a, the energy ratio of GRB/LEB is > 5. Without the GRB component, our photo-neutron yield would have been much lower. In fact our GEANT4 simulations show that the photo-neutron yield from pure hot electron bremsstrahlung gamma-rays would be ~30 times less than what is observed. We perform such GEANT4 simulations by using the observed hot electron spectrum (Fig.7) and iterate the hot electron injection number to achieve the observed MS data and dosimeter data for gamma-rays < 6 MeV. We also cross-calibrate the SAS spectrum such as Fig.1a using both photo-neutron yields (which constrain the GRB) and dosimeter data (which constrain the LEB amplitude). In general, the GRB/LEB ratio obtained from the two methods are consistent to within 20%. While this does not prove the detailed spectral shapes of either the GRB or the LEB, it serves to confirm the existence of the GRB component independent of the SAS spectrum.

On the other hand the electron and positron spectra of cm-thick targets do tightly constrain the spectral shape of the GRB, since the observed e+e- spectra faithfully represent the "birth" spectra of the pairs produced by GRB gamma-rays, with little sheath field effects or contamination from primary hot electrons. Fig.7 compares the emergent e+e- pair spectra for a cm-thick Pt target with two different GEANT4 simulations: (a) pure hot electrons injection with no additional GRB gamma-rays, (b) hot electron injection plus the GRB gamma-rays of Fig.1. We see that GEANT4 predictions for case (a) completely disagrees with observed e+e- MS data, while GEANT4 predictions for case (b) agrees well with observed MS data. It is important to note that the observed electron and positron spectra are similar in both shape and number, proving that both are dominated by pairs.

Fig.8 compares the photo-neutron yield and spectra predicted by GEANT4 for a 4.5 mm thick Pt target of case (a) and (b) above. Once again the experimental data strongly disagrees with case (a) and agrees with case (b), with a measured isotropic neutron yield of ~ $10^{10}$ neutrons.

In summary, by combining the e+e- spectral data for cm-thick targets and photo-neutron yields, we convincingly prove that the gamma-ray spectrum is completely inconsistent with hot electron bremsstrahlung alone. A large gamma-ray excess > 8 MeV above the hot electron bremsstrahlung spectrum must be present in order to produce the observed e+e- spectra and photo-neutron yields. These results independently support the GRB detected by the SAS.

## 4. THEORETICAL MODEL FOR THE ORIGIN OF THE GRB

If the TPW gamma-ray excess > 8 MeV cannot be produced by hot electron bremsstrahlung, then what could be emitting the GRB? While we are pursuing different theories, at this point the most promising and viable candidate is magnetic Compton resonant scattering (MCRS) of laser focal-spot soft photons by the relativistic hot electrons. This hypothesis is motivated by three factors: (a) the GRB is absent in directions far from laser forward; (b) the GRB peak is relatively narrow-band, (c) superstrong magnetic field is known to be present at PW laser focal spot on solid targets.

We first review the TPW laser parameters: The "50% energy focal-spot" radius of TPW is ~3 microns. The average on-target TPW energy is ~130 J. Hence the average focal-spot laser fluence is $F = 65\ \mathrm{J}/\pi(3*10^{-4}\mathrm{cm})^2 = 2.3*10^8\ \mathrm{J/cm^2}$. When a relativistic electron of Lorentz factor $\gamma >> 1$ hits a soft photon of energy $\varepsilon_o$ in the laboratory frame, the Compton formula from standard textbooks (e.g. Rybicki and Lightman 1979) gives the Compton upscattered photon energy <u>in the forward direction</u> as $\varepsilon_{sc} = 4\gamma^2\varepsilon_o$. The factor $4\gamma^2$ comes from performing the Lorentz transformation twice, from the lab. frame to the electron rest frame, and then back to the lab. frame after performing the scattering in the electron rest frame. From the Compton formula above we find that the characteristic soft photon energy $\varepsilon_o$ needed to produce the observed GRB peak is $\varepsilon_o = 12\ \mathrm{MeV}/(4\gamma^2)$ ~ 2.5 keV. However, the Thomson cross-section is tiny ($=6\mathrm{x}10^{-25}\mathrm{cm}^2$). Assuming blackbody emission at kT = 2.5 keV, the Compton scattering probability for a typical hot electron is <<< 1, whereas the observed GRB energy content is ~ 10% of observed hot electron energy. This rules out conventional Compton upscattering as a viable mechanism for the origin of the GRB.

At present our favorite mechanism for the origin of the GRB is magnetic Compton resonant scattering (MCRS, see Dermer 1990 and references therein) at the laser focal spot involving Landau resonances in a superstrong magnetic field in the electron rest frame. It is well established in many PW laser experiments that superstrong magnetic fields are created at the laser focal spot near a solid target surface (Gibbon 2005), with the maximum field limited only by the laser intensity: $B \leq B_{max} = (8\pi I/c)^{1/2}$, where I is the peak laser intensity. Even though TPW laser-created fields only reach ~few$\mathrm{x}10^9$ G in the laboratory frame, it can be boosted to ≥ few x $10^{11}$G in the hot electron rest frame, putting it at the threshold of the quantum regime (≥ few % of $B_{crit}=4\mathrm{x}10^{13}$G, similar to neutron star surface fields). Details of MCRS will be discussed in a separate paper under preparation (Liang 2024). Fig.10 illustrates the basic concept behind MCRS. The key ingredients needed for MCRS are superstrong magnetic fields, oblique hot electron pitch angles, and abundant quasi-isotropic soft photons to resonate with the cyclotron energy in the electron rest frame. Fig.11 shows a sample numerically simulated MCRS gamma-ray spectrum based on laser parameters similar to the TPW. It shows that the MCRS is capable of producing narrow-band gamma-rays peaking at ~ 10 MeV, with amplitude up to $10^5$ times higher than non-magnetic Compton scattering. The peak gamma-rays are narrowly beamed along the hot electron direction. For a given soft photon distribution, only hot electrons of appropriate energies can resonate with them, thus producing narrow-band gamma-rays whose peak energy varies with angle. The predicted MCRS gamma-ray peak energy and intensity scale as high powers of the laser intensity and laser energy. Hence we can potentially test the MCRS theory by varying the laser energy and laser intensity, and measuring the gamma-ray spectrum versus angle from laser forward. This will be pursued in future PW laser experiments.

In summary, Compton scattering of soft photons by hot electrons may be enhanced by many orders of magnitude above the Thomson value in the presence of superstrong magnetic fields at the laser focal spot near a solid target surface.

ACKNOWLEGEMENTS

At Rice University this work was supported by DOE grant DE-SC0021327 and DE-SC0024874. At UT Austin, this work was supported by the DOE, Office of Science, Fusion Energy Sciences under Contract No. DE-SC0021125: LaserNetUS: A Proposal to Advance North America's First High Intensity Laser Research Network".

Table 1. TPW laser parameters of our 2022 60-shot run

| | Energy | Pulse Duration (fs) | Peak Power (TW) | closed Radius with 50% (u | ak Intensity (W/cm | Strehl |
|---|---|---|---|---|---|---|
| AVERAGE | 123.25 | 161.05 | 781.82 | 3.8 | 2.96E+21 | 0.68 |
| MEDIAN | 122.57 | 158 | 794.12 | 3.72 | 2.89E+21 | 0.7 |
| MIN | 104.2 | 128 | 531.29 | 2.42 | 1.85E+21 | 0.46 |
| MAX | 139.18 | 216 | 1060.68 | 5.39 | 4.68E+21 | 0.82 |

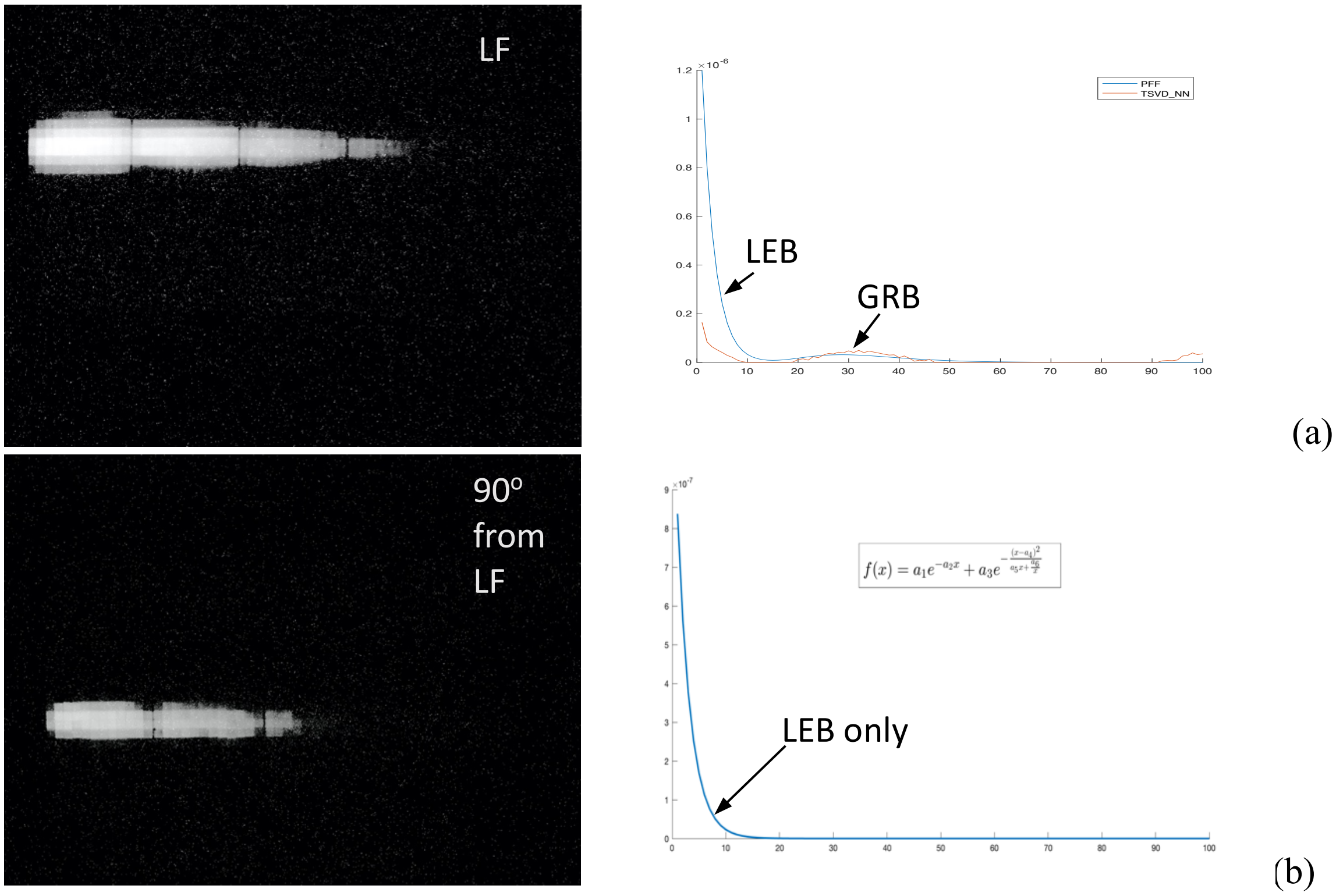

Fig.1 (a) Sample raw SAS image (left) and inverted gamma-ray spectrum (right) obtained at laser forward (LF) from our 2022 TPW experiment. TSVDNN denotes the model-independent spectrum directly inverted from the image. PFF denotes the best-fit analytic spectrum derived from the TSVDNN solution. Both spectra show two distinct spectral components. LEB denotes the low-energy bremsstrahlung component below 8 MeV and GRB denotes the broad high-energy bump above 8 MeV. See Liang et al 2022 for discussions of our inversion algorithm; (b) sample SAS image (left) and inverted spectrum. (right) obtained at 90° from LF. Here the best-fit spectrum shows no GRB component. The x-axes of both spectra are in units of 0.5 MeV. The y-axes units are arbitrary. Almost all SAS spectra obtained at LF and 90° from LF resemble these two examples.

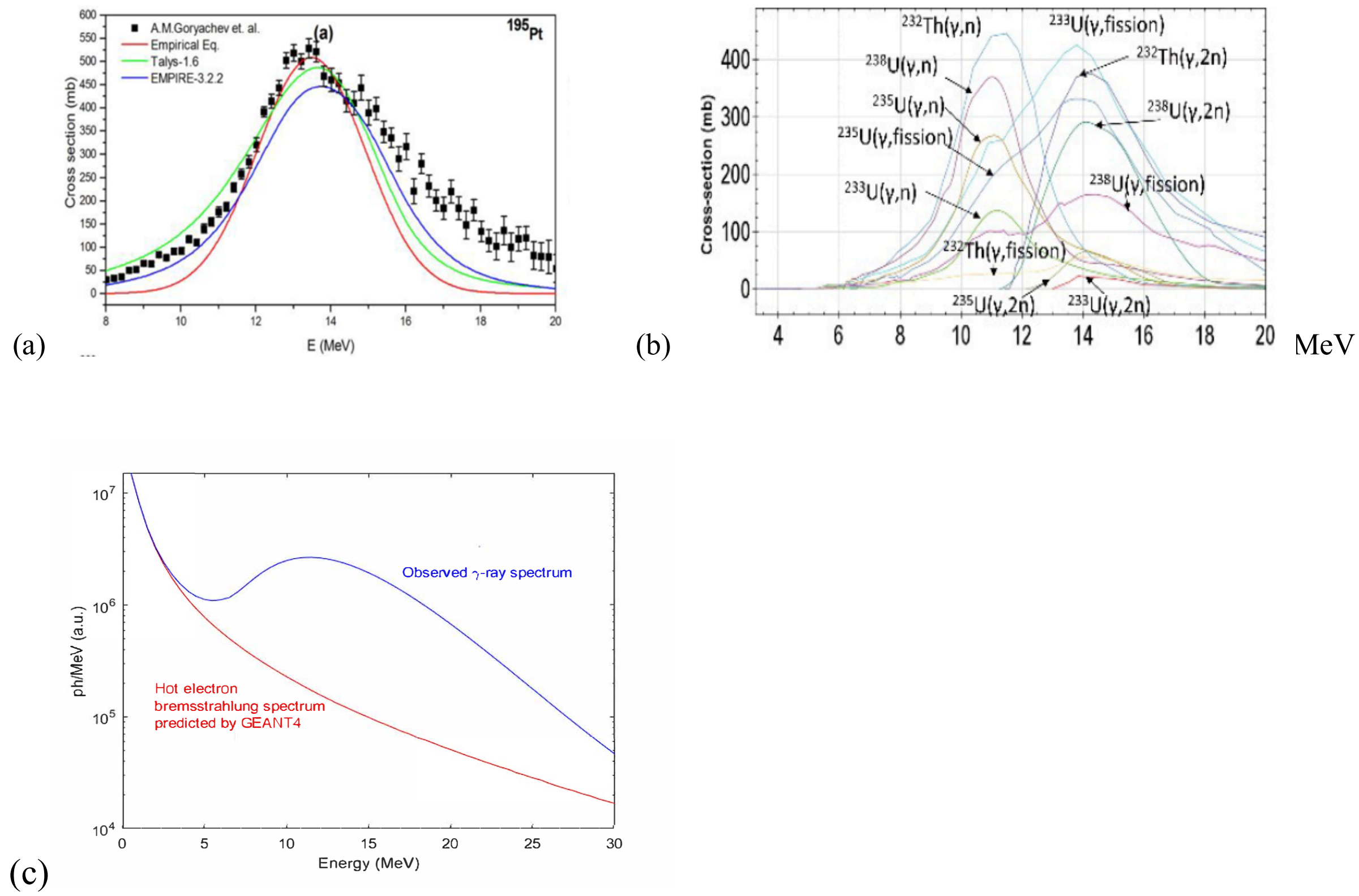

Fig.2 (a) (γ, n) photo-neutron reaction cross-section of Pt shows the GDR peak at ~ 13.5 MeV (from Makwana et al 2017); (b) (γ, Nn) and (γ, fission) reaction cross-sections of Th and U isotopes show GDR peaks between 11 and 16 MeV (from Kimura et al 2016). All GDR peaks of high-Z elements span ~ 8 MeV to ~20 MeV, (c) Replot of Fig.1(a) blue curve in log-linear scale compared to GEANT4 simulated hot electron bremsstrahlung spectrum to highlight the GRB excess > 8 MeV. The GRB strongly overlaps the GDR peaks of Fig.2(a) and Fig.2(b).

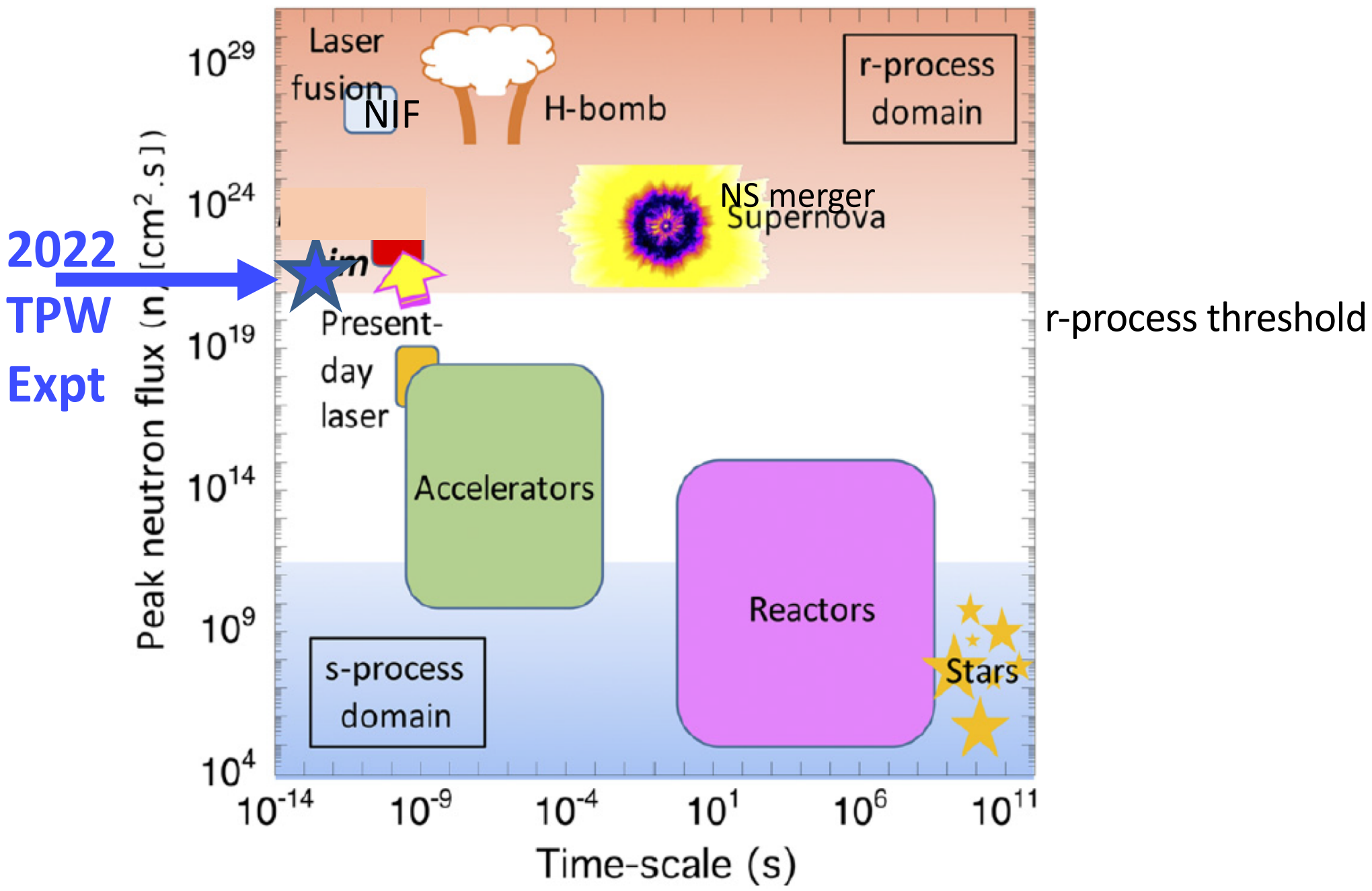

Fig.3 Comparison of peak neutron flux achieved in our 2022 TPW experiment with other man-made and cosmic neutron fluxes relevant to r-process element creation (figure adapted from Chen et al 2019).

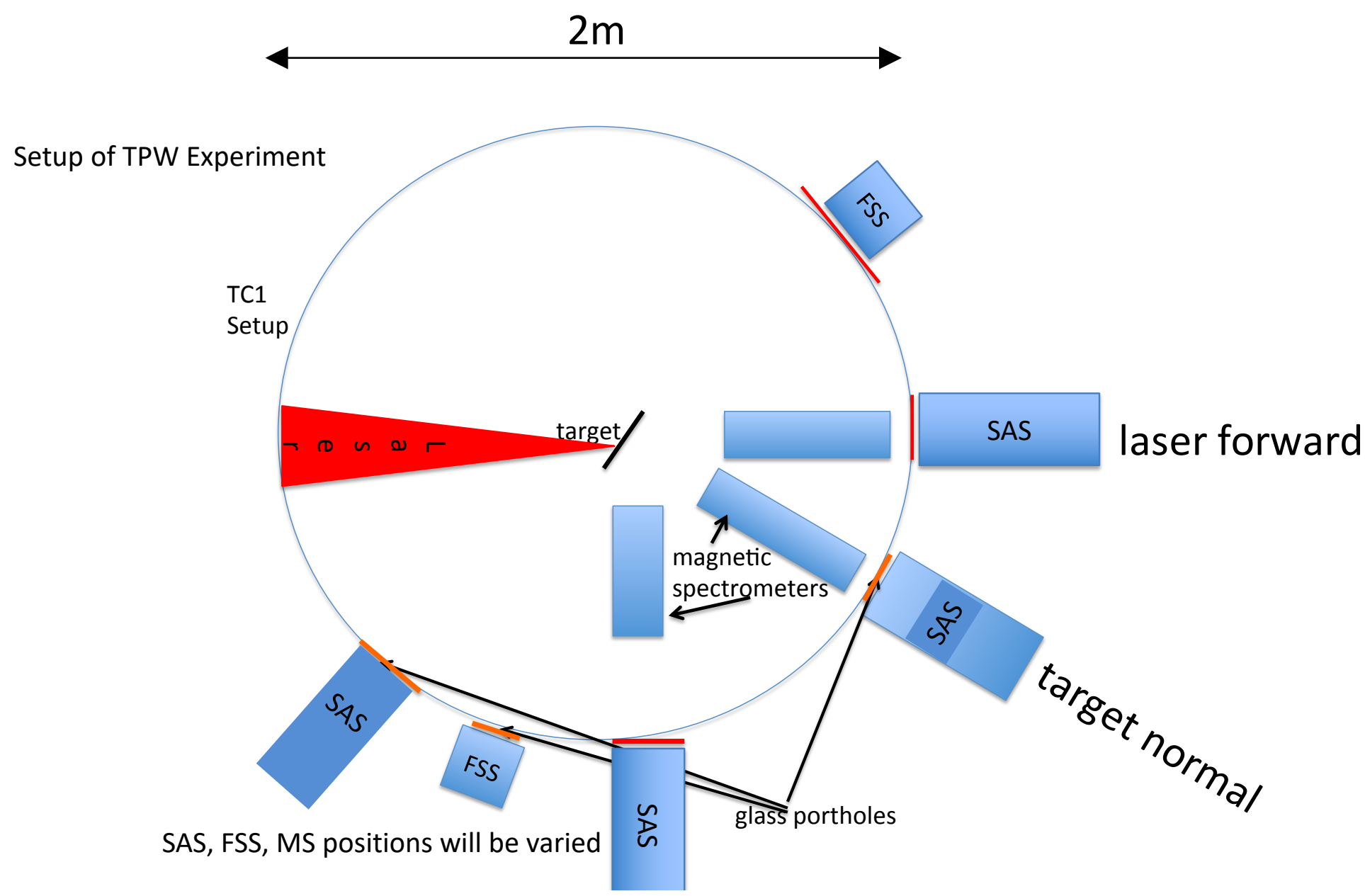

Fig.4 Typical diagnostics set-up in TC1 of TPW. Not shown are the gamma dosimeters and neutron bubble detectors attached to the outside of the tank wall. To minimize attenuation most FSS's and SAS's view the target through aluminum windows. In 2022 the laser incidence angle was 17°. The target orientation was s-polarized.

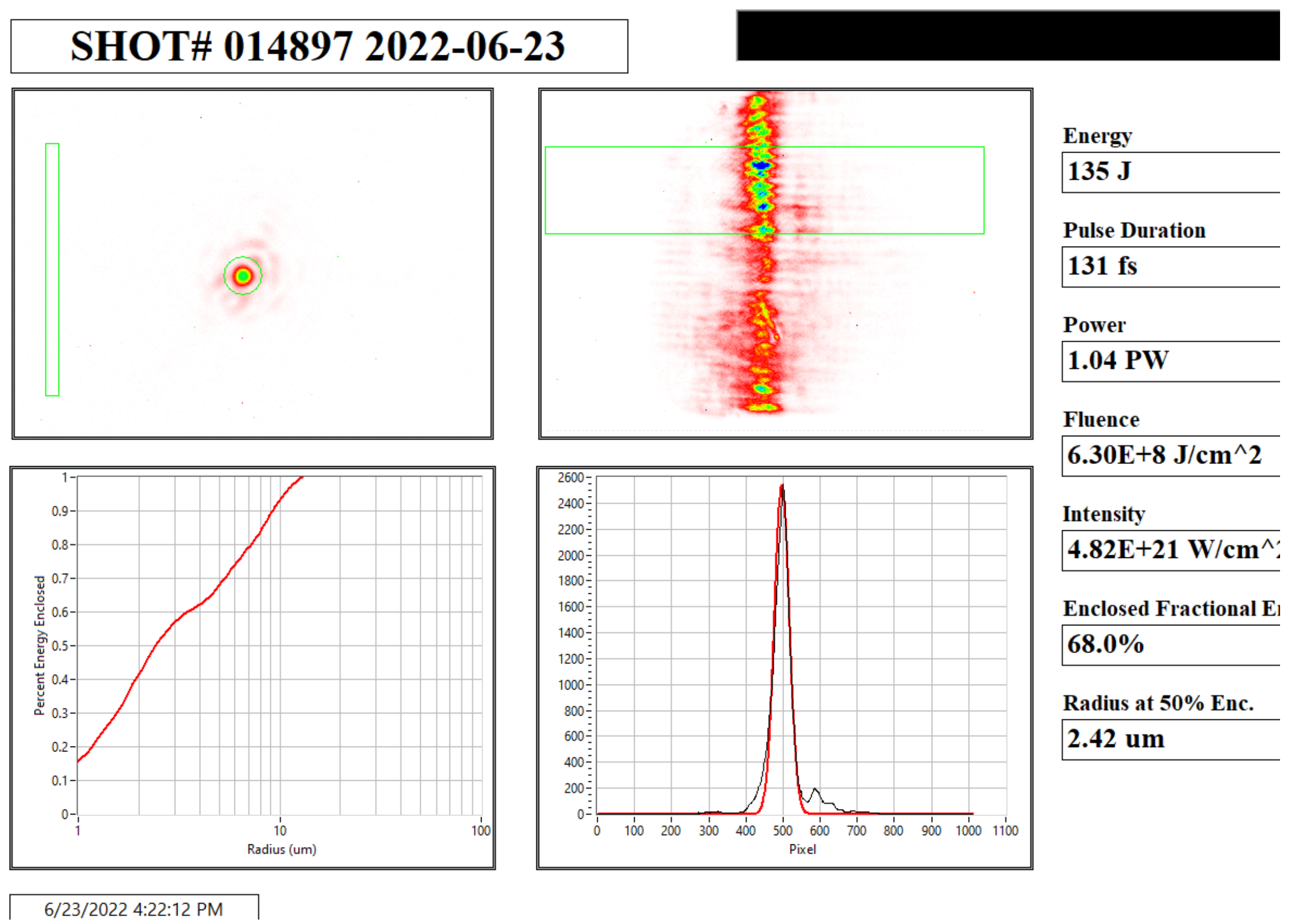

Fig.5 Sample laser pulse spatial and temporal profiles of a TPW shot during our 2022 run. The peak intensity was 4.8x $10^{21}$W.cm$^{-2}$ for this shot.

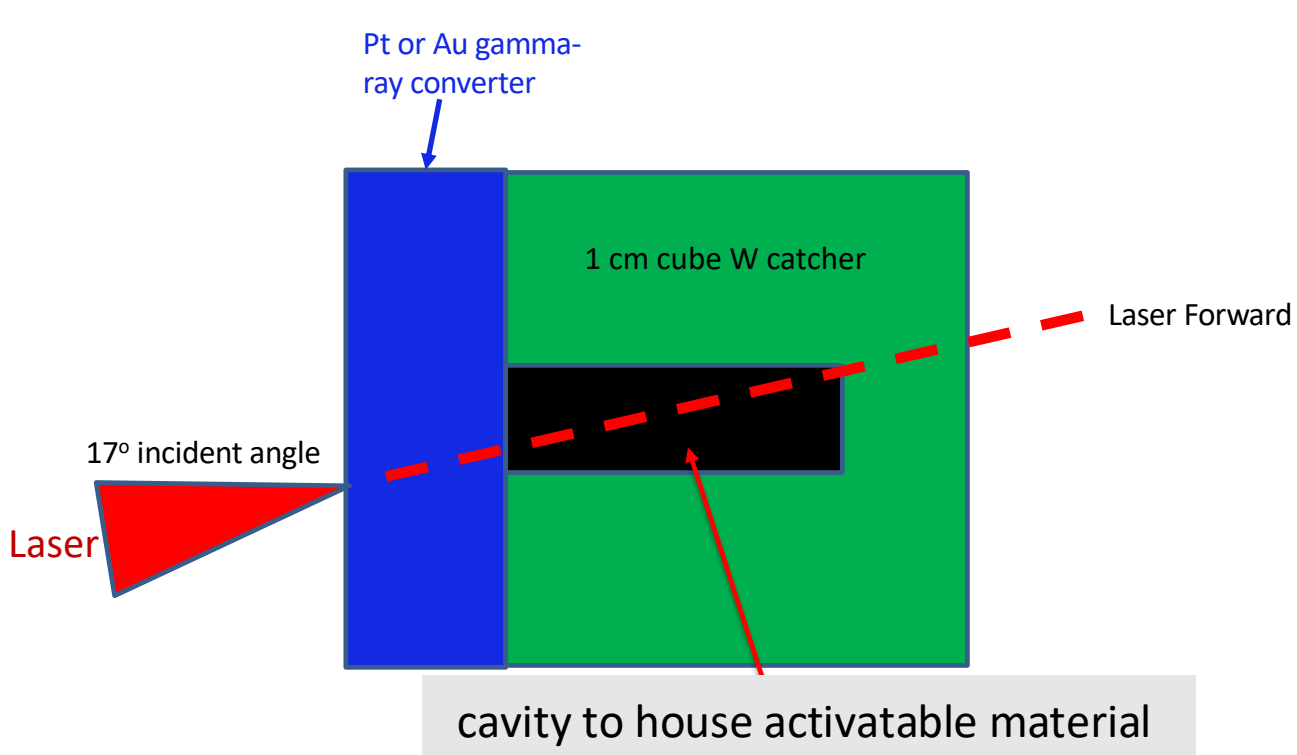

Fig.6 Sketch of capsule target design to fully contain and shield any potentially radioactive products from post-shot targets for safe and easy handling. The capsule consists of a 1 cm cube of W catcher with a cylindrical divot or cavity to house any activatable material. The Pt or Au gamma-ray converter serves as a removable capsule cover. This design worked well in our 2022 experiment. The laser aiming position on the front surface varies with converter thickness such that the forward gamma-ray cone covers the desired volume of the cavity. Due to the width of the gamma-ray cone we can shoot the same target multiple times by simply shifting the target position by a small amount, depending on the crater size of the previous shot. In 2022, we managed to shoot the same target up to 5 times. All data reported in this paper came from flat disks and empty capsules only.

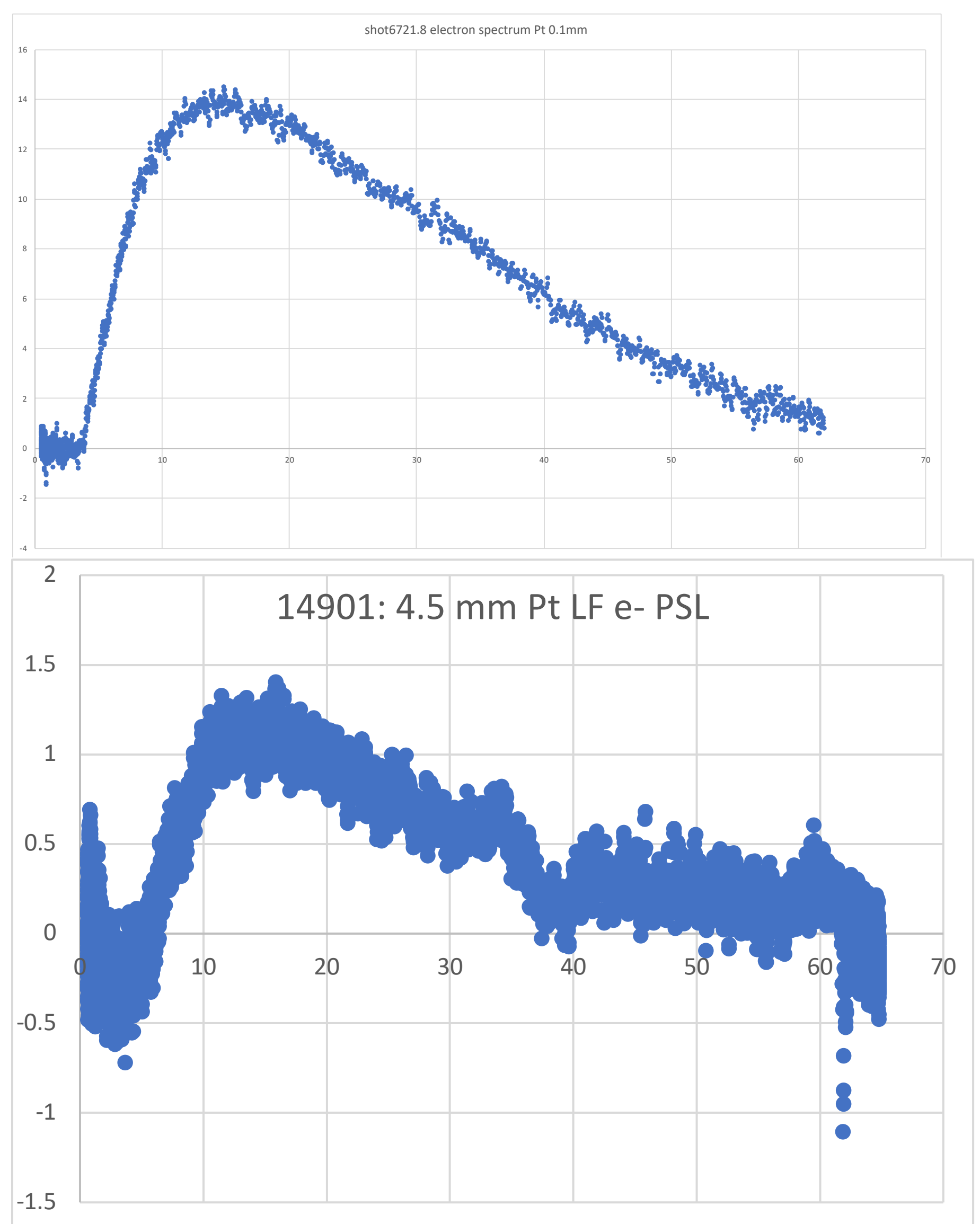

Fig.7 Sample measured hot electron spectra from a 0.1 mm thick Pt target (top) and a 4.5 mm thick Pt target (bottom) show similar spectral shapes despite a factor of 45 difference in target thickness. The relative stability of the hot electron spectral shape and peak energy vs. target thickness allows us to confidently use this spectral shape for hot electron injection in GEANT4 simulations. We scale the electron injection number to match the predicted emergent outputs in e+e- pairs, gamma-rays and neutrons with experimental data, and then check for consistency among the different data sets.

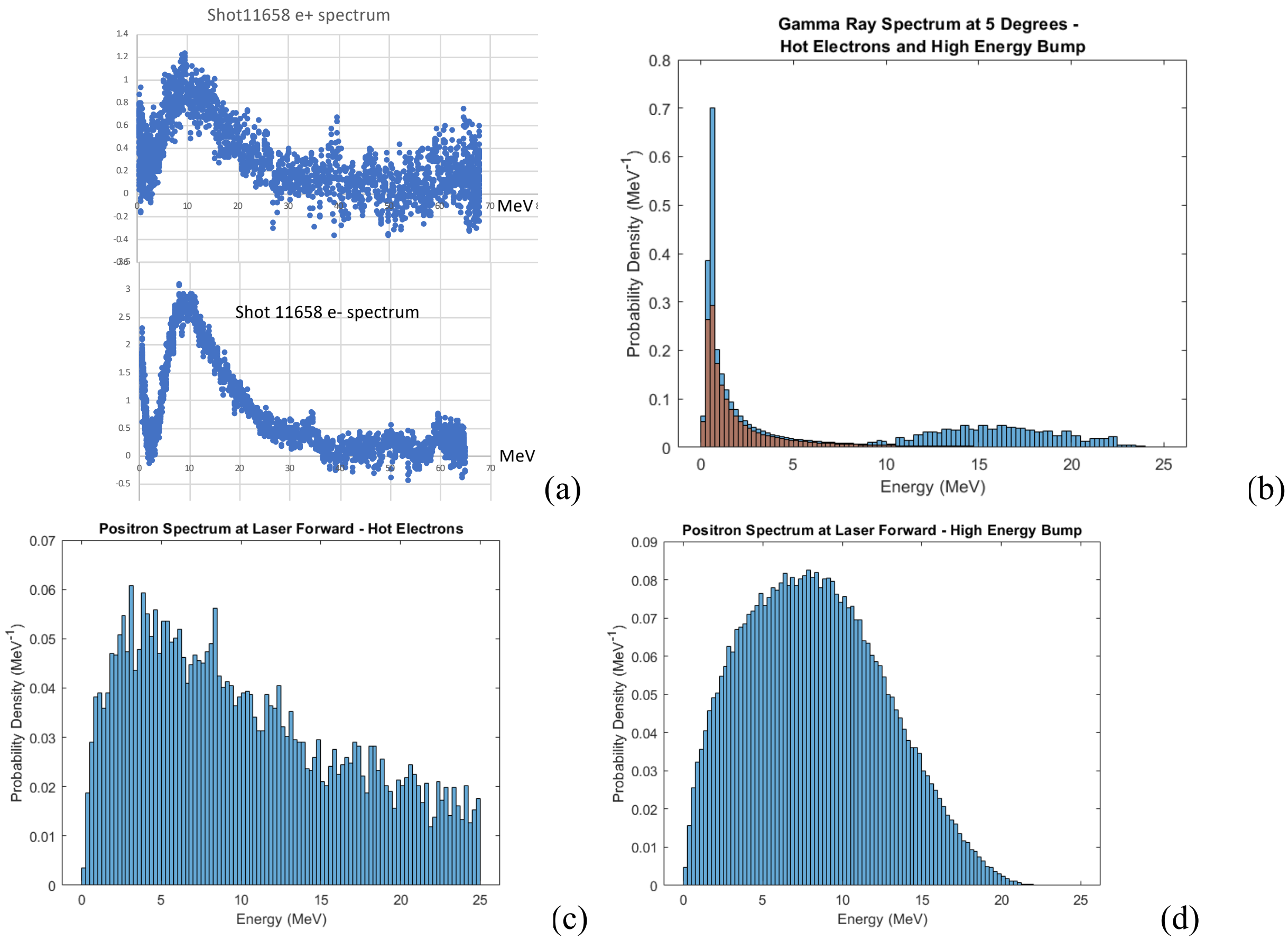

Fig.8 (a) Observed electron and positron spectra for a thick Pt target show almost identical shapes, peak energies and amplitude, showing that both are created by the same gamma-rays with no sheath field effects. The positron spectrum peaks at ~ 8-9 MeV. All primary hot electrons are absorbed in such thick targets except for the high energy tail above 30 MeV. (b) GEANT4-predicted gamma-rays from two different injection models: brown curve: only hot electrons of Fig.7 spectrum without additional gamma-rays from the GRB of Fig.1; blue curve: hot electrons plus GRB gamma-rays of Fig.1, (c) predicted positron spectrum from pure hot electron injection without the GRB gamma-rays disagrees with the positron spectrum observed in Fig.8(a). The positrons created by bremsstrahlung gamma-rays peak at ~2-3 MeV. (d) predicted positron spectrum from injection of hot electrons plus GRB gamma-rays peak at ~ 8 MeV, which roughly agrees with the observed positron spectrum in Fig.8(a).

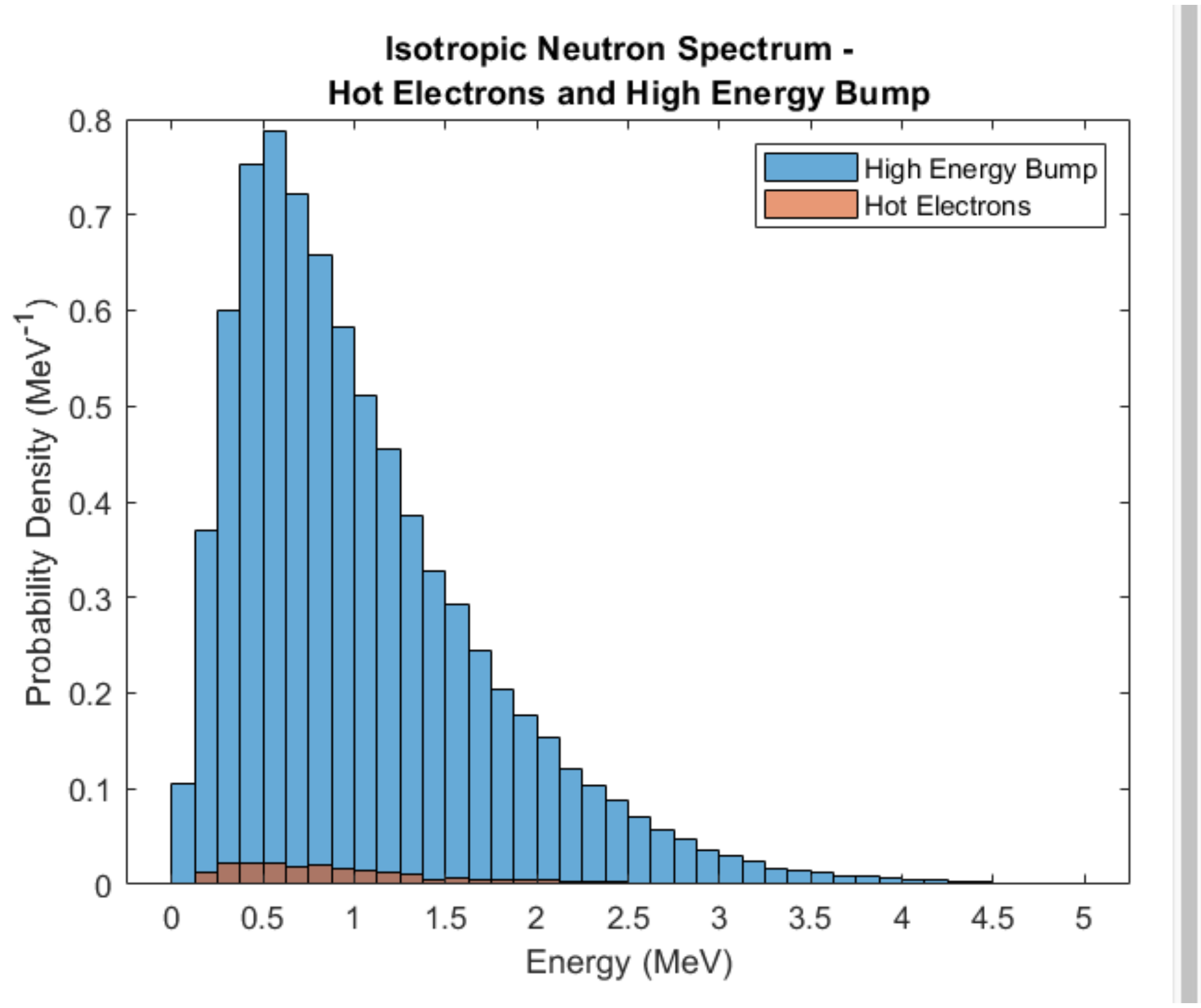

Fig.9 GEANT4-predicted emergent neutron spectra from the two types of injection models used in Fig.8(b): brown curve: hot electrons only without GRB gamma-rays of Fig.1a; blue curve: hot electrons plus GRB gamma-rays of Fig.1a. The predicted neutron number of the blue curve agrees with bubble detector data to within ~ 20%, whereas the predicted neutron number of the brown curve is ~ 30 times lower than observed.

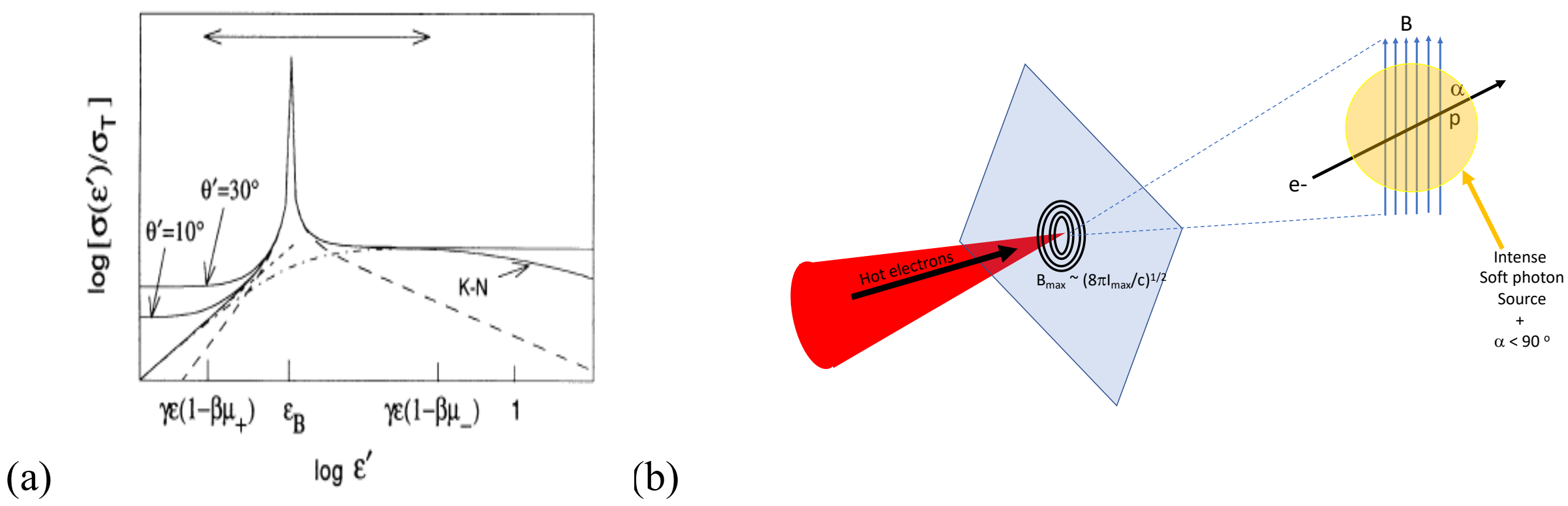

Fig.10 (a) Compton cross-section in a superstrong magnetic field exhibits a sharp resonance at the cyclotron energy $\varepsilon_B$(=heB/2πmc) in the electron rest frame, while the cross-section is suppressed below $\varepsilon_B$. The amplitude of the resonance can be many orders of magnitude higher than the Thomson cross-section. Even though the resonance is narrow, soft photons of a wide energy range can resonate due to the Doppler effect (from Dermer 1990), (b) Blowup picture of PW laser focal spot magnetic field together with a sample hot electron track and soft photon source. Hot electron must have oblique pitch angle ($\alpha < 90^o$) for MCRS to occur (adapted from Liang 2024).

Magnetic Resonant Scattering produces narrow-band beamed gamma-rays many orders of magnitude higher than Non-magnetic Compton Scattering

*$B_o$=5x$10^9$ G, $\omega_B$= 60 eV, pitch angle = 70$^o$, $f_e(E)$ ~ exp( -E/20 MeV)*

Mag Res Scatt Spectrum

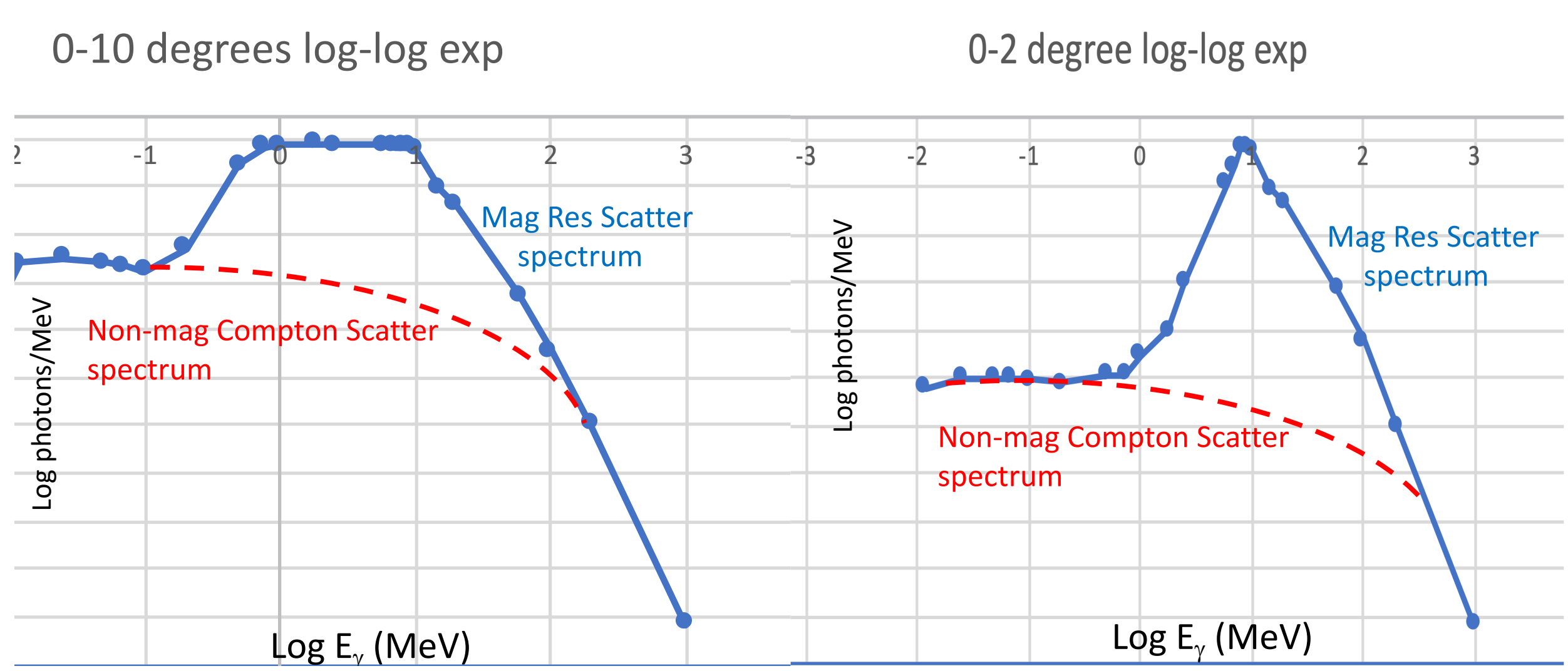

Fig.11 Numerically simulated MCRS spectra (blue) for laser parameters similar to the TPW. Input values of magnetic field, soft photon energy, hot electron spectrum and pitch angle are listed at the top of the figure. Red dashed curves represent Compton upscattered spectra without the magnetic resonance, which lie many orders of magnitude below the blue curves. Note that the MCRS spectral peak broadens as wider angles are included (adapted from Liang 2024).